\documentclass[12pt,prd,aps,amssymb,amsmath,tightenlines,showpacs]{article}
\usepackage[utf8]{inputenc}
\usepackage[a4paper,top=3cm,bottom=3cm,left=1.5cm,right=1.5cm]{geometry}
\usepackage{amssymb}
\usepackage{amsmath}
\usepackage{amsthm}
\usepackage{physics}
\usepackage{mathtools}
\usepackage{tipa}
\usepackage{latexsym} 
\usepackage{graphicx}
\usepackage{slashed}
\usepackage{bbold}
\usepackage{cancel}
\usepackage{comment}
\usepackage{color}
\usepackage{graphicx,epsfig,color}
\usepackage[dvipsnames]{xcolor}
\usepackage{comment}
\usepackage{cite}
\usepackage{tikz}
\usepackage[compat=1.1.0]{tikz-feynman}
\usepackage{caption}
\usepackage{subcaption}

\newcommand{\be}{\begin{equation}}
	\newcommand{\ee}{\end{equation}}
\newcommand{\bea}{\begin{eqnarray}}
	\newcommand{\eea}{\end{eqnarray}}
\newcommand{\vv}{``}

\newcommand{\mpl}{M_P}

\begin{document}
	\graphicspath{{FIGURE/}}
	\topmargin=-1cm
	
	\begin{center} 
	{\Large{\bf Path integral measure and cosmological constant}}\\		
		
		\vspace*{0.8 cm}
		
		Carlo Branchina\footnote{carlo.branchina@unical.it}\label{one}$^{\dagger}$,
		Vincenzo Branchina\footnote{branchina@ct.infn.it}\label{two}$^\ddagger$, 
		Filippo
		Contino\footnote{filippo.contino@ct.infn.it}\label{three}$^\ddagger$,
		Arcangelo
		Pernace\footnote{arcangelo.pernace@ct.infn.it}\label{four}$^\ddagger$
		\vspace*{0.4cm}
		
	   {\it			
			${}^\dagger$Department of Physics, University of Calabria, and INFN-Cosenza\\
			Arcavacata di Rende, I-87036, Cosenza, Italy 
			
			\vskip 5pt
			
			${}^\ddagger$Department of Physics, University of Catania, and INFN-Catania\\
			Via Santa Sofia 64, I-95123 
			Catania, Italy
		}
			
	 \vskip 20pt
	 {\bf Abstract}
	 \noindent
	 
\end{center}

{\small 
\noindent
Considering (euclidean) quantum gravity in the Einstein-Hilbert truncation, we calculate the one-loop effective action $\Gamma^{1l}_{\rm grav}$ using a spherical background. 
Usually, this calculation is performed resorting to proper-time regularization within the heat kernel expansion and gives rise to quartically and quadratically UV-sensitive contributions to the vacuum energy $\rho_{\rm vac}=\frac{\Lambda_{\rm cc}}{8\pi G}$, with $\Lambda_{\rm cc}$ and $G$ cosmological and Newton constant, respectively. We show that, if the measure in the path integral that defines $\Gamma^{1l}_{\rm grav}$ is correctly taken into account, and the physical UV cutoff $\Lambda_{\rm cut}$  properly introduced, $\rho_{\rm vac}$ presents only a (mild) logarithmic sensitivity \,to\, $\Lambda_{\rm cut}$. We also consider a free scalar field and a free Dirac field on a spherical gravitational background, and find that the same holds true even in the presence of matter. These results are found without resorting to any supersymmetric embedding of the theory, and shed new light on the cosmological constant problem.}

\section{Introduction}

It has been known for a long time that our universe is going through a phase of accelerated expansion \cite{perl}, a feature that might be explained with the introduction of a \vv tiny", positive, vacuum energy  $\rho_{\rm vac}\sim2.7\cdot10^{-47}\,\text{GeV}^4$ \cite{ParticleDataGroup:2024cfk} in Einstein's equations. Field theory calculations, typically performed using proper-time regularization within the heat kernel expansion \cite{Schwinger:1951nm,DeWitt:1964mxt,Seeley1,Seeley2,DeWitt:1975ys} (see for instance\,\cite{FradkinTseytlin,TaylorVeneziano}), show the appearance of quartically and quadratically UV-sensitive terms (\vv divergences") in the perturbative evaluation of the vacuum energy. If the UV physical cutoff of a gravitational field theory is identified with the Planck scale $\mpl$, the quartically divergent contribution to $\rho_{\rm vac}$ is $\sim\mpl^4$, and the comparison with the experimental value reported above shows a discrepancy of approximately $123$ orders of magnitude, that calls for an enormous and quite unnatural fine-tuning. Clearly, when calculations are performed with dimensional regularization (DR), power divergences are absent since they are cancelled by construction, though they are somehow encoded in the presence of poles in dimensions lower than four (see \cite{Branchina:2022jqc} for a recent and thorough study of DR). In this respect, DR only
realizes a technical cancellation, that does not provide any information on the physical mechanism responsible for the elimination of the strongly UV-sensitive contributions from the highest momentum modes.

The most studied and well-known way to realize a physical cancellation of power-like UV-sensitive terms consists in performing a supersymmetric embedding of the theory (supergravity). The latest LHC runs up to TeV scales, however, have not shown any evidence for supersymmetry (SUSY). If present, SUSY must be broken at higher energies. Therefore, though supergravity successfully cancels the $\mpl^4$ and $\mpl^2$ terms, contributions to $\rho_{\rm vac}$ proportional to the fourth power of particles masses (and/or vevs) are left. If the SUSY breaking scale $M_{\rm SUSY}$ is not hierarchically higher than the Fermi scale, the problem is sensibly alleviated, although still present at the level of about $50$ orders of magnitude. A complete solution to the cosmological constant problem necessarily requires a mechanism that not only disposes of the large, power-like UV-sensitive terms, but also cancels the $({\rm mass})^4$ contributions, driving the vacuum energy $\rho_{\rm vac}$ down to its measured value. There have also been attempts to implement the cancellation of power-like divergences considering models with compact extra dimensions, but it has been recently shown that terms undoing the cancellation were missed (see \cite{Branchina:2023ogv, Branchina:2023rgi, Branchina:2024ljd} and references therein)

In this work we re-examine the calculation of the one-loop effective action $\Gamma^{1l}_{\rm grav}$ in (euclidean) quantum gravity. We show that, differently from what is typically acknowledged, the quantum dressing of $\rho_{\rm vac}$ does not produce neither quartic nor quadratic divergences, regardless of any symmetry that may be added to (incorporated in) the theory. Moreover, we show that the appearance of power-like divergences in usual calculations is due to an incorrect treatment of the measure in the path integral that defines $\Gamma^{1l}_{\rm grav}$ and to an improper implementation of the physical UV cutoff $\Lambda_{\rm cut}$. Once the measure and the physical UV cutoff are properly treated, $\rho_{\rm vac}$ turns out to be only logarithmically sensitive to the UV scale\footnote{Concerning quartic divergences, their absence for a massless scalar theory on a gravitational background was first pointed out in \cite{Fradkin:1976xa}. For pure gravity, the absence of these quartic divergences was already noted in\,\cite{Donoghue:2020hoh}.}. Though this does not provide a complete solution to the cosmological constant problem, these results show that the latter is less severe than typically acknowledged.

The rest of the paper is organized as follows. Considering pure gravity in the Einstein-Hilbert truncation, in section \ref{oneloop pure gravity} we calculate the one-loop (euclidean) effective action using a spherical background. 
In section \ref{oneloop}, we discuss our results for the one-loop cosmological constant $\Lambda_{\rm cc}$ and the Newton constant $G$, and compare them with those of previous literature. In section \ref{matter}, we consider a free massive scalar field $\phi$ (section \ref{scalar}) and a free massive Dirac field $\psi$ (section \ref{fermion}) on a spherical gravitational background, and calculate the dressing of the Newton and cosmological constant due to the quantum fluctuations of these fields. Section \ref{conclusions} is for conclusions.

\section{One-loop effective action. Pure gravity}
\label{oneloop pure gravity}
Let us consider the (euclidean) gravitational action in the Einstein-Hilbert truncation
\begin{equation}
	S_{\rm grav}=\frac{1}{16\pi G}\int\dd[4]x\,\sqrt{g}\,\left(-R+2\Lambda_{\rm cc}\right)\,.
	\label{bareaction}
\end{equation}
To calculate the one-loop effective action $\Gamma^{1l}_{\rm grav}=S_{\rm grav}+\delta S^{1l}_{\rm grav}$, we resort to the geometrical approach pioneered by Vilkovisky\,\cite{Vilkovisky:1984st} and DeWitt\,\cite{DeWitt:1987te} that allows to obtain a gauge invariant result even  off-shell. We follow the strategy put forward in\,\cite{FradkinTseytlin, TaylorVeneziano}, paying particular attention to the measure in the path integral that defines $\Gamma^{1l}_{\rm grav}$. Considering the background field method\,\cite{Abbott:1980hw,Abbott:1981ke}, we write the metric $g_{\mu\nu}$ as $g_{\mu\nu}=\Bar g_{\mu\nu}+h_{\mu\nu}$, where $\Bar g_{\mu\nu}$ is the background and $h_{\mu\nu}$ the fluctuation. As shown in \cite{FradkinTseytlin}, when $\Bar g_{\mu\nu}$ has spherical symmetry the one-loop Vilkovisky-DeWitt effective action coincides with the standard one calculated with gauge-fixing
\begin{equation}\label{gf}
	S_{\rm gf}=\frac{1}{32\pi G\xi}\int\dd[4]{x}\sqrt{\Bar g}\left[\nabla_\mu\left(h^\mu_{\nu}-\frac12\delta^\mu_{\nu}\,h^{\sigma}_{\sigma}\right)\right]\,
\end{equation}
taking the limit $\xi\to0$ at the end of the calculation. We then consider a spherical background $\Bar g_{\mu\nu}=g^{(a)}_{\mu\nu}$ ($a$ radius of the sphere), and choose the coordinates $x$ to be dimensionless so that $g^{(a)}_{\mu\nu}$ is proportional to $a^2$. For this background, the classical action\,\eqref{bareaction} is ({\small $\int\dd[4]x\,\text{\footnotesize$\sqrt{g^{(a)}}$}=\frac{8\pi^2}{3}a^4$\,,\, $R(g^{(a)})=\frac{12}{a^2}$})
\begin{equation}
	S^{(a)}_{\rm grav}\equiv S_{\rm grav}[g^{(a)}_{\mu\nu}]=\frac{\pi\Lambda_{\rm cc}}{3G}a^4-\frac{2\pi}{G}a^2\,,
	\label{Sb}
\end{equation}
and the ghost action $S_{\rm ghost}$ corresponding to $S_{\rm gf}$ in\,\eqref{gf} is
\begin{equation}\label{ghostaction}
	S_{\rm ghost}=\frac{1}{32\pi G}\int\dd[4]{x}\sqrt{g^{(a)}}\,g^{(a)\,\mu\nu}\,v_\mu^*\left(-\nabla_\rho\nabla^{\rho}-\frac{3}{a^2}\right)v_\nu\,.
\end{equation}
After the calculation of $\delta S^{1l}_{\rm grav}$, we will identify the one-loop corrections $\delta\left(\frac{\Lambda_{\rm cc}}{G}\right)$ and $\delta\left(\frac{1}{G}\right)$ to $\frac{\Lambda_{\rm cc}}{G}$ and $\frac{1}{G}$ with the coefficients of $a^4$ and $a^2$ respectively \cite{TaylorVeneziano}. 

According to the strategy outlined above, the one-loop correction $\delta S^{1l}_{\rm grav}$ is obtained from 
\begin{align}\label{effacgrav1}
	e^{-\delta S^{1l}_{\rm grav}}=\lim_{\xi\to0}\int\big[\mathcal{D}u(h)\mathcal{D}v_\rho^*\,\mathcal{D}v_\sigma\big]\, e^{-\delta S^{(2)}}\,,
\end{align}
where 
\begin{equation}\label{deltaS2}
\delta S^{(2)}\equiv S_2+S_{\rm gf}+S_{\rm ghost}\,,
\end{equation}
with $S_2$ the quadratic term in the expansion of $S_{\rm grav}[g^{(a)}_{\mu\nu}+h_{\mu\nu}]$ around $g^{(a)}_{\mu\nu}$ (below {\small $h\equiv g^{(a)}_{\mu\nu}h^{\mu\nu}$\,,\, $\tilde h_{\mu\nu}\equiv h_{\mu\nu}-\frac12 g^{(a)}_{\mu\nu} h$}\,; indexes are raised with {\small $g^{(a)\,\mu\nu}$}, and covariant derivatives are in terms of {\small $g^{(a)}_{\mu\nu}$})
\begin{equation}\label{S2}
	S_2\equiv\frac{1}{32\pi G}\int\dd[4]x\,\sqrt{g^{(a)}}\left[\frac12\tilde h^{\mu\nu}\left(-\nabla_\rho\nabla^{\rho}-2\Lambda_{\rm cc}+\frac{8}{a^2}\right)h_{\mu\nu}+\frac{h^2}{a^2}-\nabla^{\rho}\tilde h_{\rho\mu}\nabla^{\sigma}\tilde h_{\sigma}^{\mu}\right]\,.
\end{equation}
The measure $\big[\mathcal{D}u(h)\mathcal{D}v_\rho^*\,\mathcal{D}v_\sigma\big]$ is given by
\begin{equation}\label{meas}
	\big[\mathcal{D}u(h)\mathcal{D}v_\rho^*\,\mathcal{D}v_\sigma\big]\equiv \prod_x\Big[g^{(a)\,00}(x)\,\left(g^{(a)}(x)\right)^{-1}\Big(\prod_{\alpha\leq\,\beta}\dd{h_{\alpha\beta}(x)}\Big)\Big(\prod_{\rho}\dd{v_\rho^*(x)}\Big)\Big(\prod_{\sigma}\dd{v_\sigma(x)}\Big)\Big]\,.
\end{equation}
In \cite{Fradkin:1973wke}, Fradkin and Vilkovisky show that the terms {\small $g^{(a)\,00}(x)\,\left(g^{(a)}(x)\right)^{-1}$} come from the integration over conjugate momenta\footnote{The alert reader might notice that the original expression in\,\cite{Fradkin:1973wke} is $g^{(a)\,00}(x)\,\left(g^{(a)}(x)\right)^{-\frac32}$. The difference is due to the fact that here we take both $v$ and $v^*$ as world vectors, while in\,\cite{Fradkin:1973wke} a different choice is made.
The extra factor $\sqrt{g^{(a)}}$ is the Jacobian that relates these two equivalent functional integration variables\,\cite{Unz:1985wq}.}, and that the measure\,\eqref{meas} is invariant under general coordinate transformations. We will see that these terms play a crucial role in the calculation of $\delta S^{1l}_{\rm grav}$. We also note that, following \cite{Honerkamp:1971xtx}, an equivalent expression of the measure was given in \cite{Donoghue:2020hoh}. 

We now observe that the background metric $g^{(a)}_{\mu\nu}$ can be written as 
\begin{equation}\label{metric1}
	g^{(a)}_{\mu\nu}=a^2 \,\widetilde g_{\mu\nu}\,,
\end{equation}
where the elements of\, $\widetilde g_{\mu\nu}$ are dimensionless and $a$-independent. The factor {\small $g^{(a)\,00}(x)$ $\left(g^{(a)}(x)\right)^{-1}$} that appears in the measure\,\eqref{meas} can then be written as
\begin{equation}\label{measterms}
g^{(a)\,00}(x)\,\left(g^{(a)}(x)\right)^{-1}=a^{-10}\,\widetilde g^{\,00}(x)\,\Big(\widetilde g(x)\Big)^{-1}\,.
\end{equation}
From the above equation we see that the $a$-dependence of the non-trivial terms in the measure is all contained in the factor $a^{-10}$.
As we will see below, it is convenient to define the dimensionless field
\begin{equation}\label{hath}
	\widehat h_{\mu\nu}\equiv\left(32\pi G\right)^{-1/2}a^{-1}h_{\mu\nu}\,,
\end{equation}
so that $S_2+S_{\rm gf}$ can be written as
{\begin{equation}\label{S2+Sgf}
		S_2+S_{\rm gf}=\int\dd[4]x\,\sqrt{\,\widetilde g\,\,}\left[\frac12\overline h^{\mu\nu}\left(-\nabla_\rho\nabla^{\rho}-2a^2\Lambda_{\rm cc}+8\right)\widehat h_{\mu\nu}+\widehat h^2-\left(1-\frac{1}{\xi}\right) (\nabla^{\rho}\overline h_{\rho\mu})\,(\nabla^{\sigma}\overline h_{\sigma}^{\mu}\,)\right]\,,
\end{equation}}

\noindent
where $\widehat h\equiv \widetilde g_{\mu\nu}\,\widehat h^{\mu\nu}$\,, $\overline h_{\mu\nu}\equiv \widehat h_{\mu\nu}-\frac12\,\widetilde g_{\mu\nu}\,\widehat h$, indexes are raised with $\widetilde g^{\,\mu\nu}$ ($\nabla^\rho\equiv\widetilde g^{\,\rho\sigma}\,\nabla_\sigma$), and covariant derivatives are defined in terms of $\widetilde g_{\mu\nu}$. 
Note also that in\,\eqref{S2+Sgf} only dimensionless operators, namely $-\nabla_\rho\nabla^{\rho}$ and $-\nabla^\rho$, appear. Moreover, introducing\,\eqref{metric1} in\,\eqref{ghostaction}, and defining the dimensionless field
\begin{equation}\label{hat v}
\widehat v_\mu\equiv\left(32\pi G\right)^{-\frac12}v_\mu\,,
\end{equation}
$S_{\rm ghost}$ can be written as (again indexes are raised with $\widetilde g^{\,\mu\nu}$ and covariant derivatives are in terms of $\widetilde g_{\mu\nu}$)
\begin{equation}\label{ghostaction2}
	S_{\rm ghost}=\int\dd[4]{x}\sqrt{\,\widetilde g\,\,}\,\,\widetilde g^{\,\mu\nu}\,\widehat v_\mu^{\,*}\left(- \,\nabla_\rho\nabla^{\rho}-3\right)\widehat v_\nu\,.
\end{equation}
Finally, we observe that
\begin{equation}\label{prodh}
	\prod_{\alpha\leq\,\beta}\dd{h_{\alpha\beta}(x)}=\left
	(32\pi G\right)^5 a^{10}\prod_{\alpha\leq\,\beta}\dd{\widehat h_{\alpha\beta}(x)}\,.
\end{equation}
Using\,\eqref{measterms},\,\eqref{hat v} and\,\eqref{prodh}, the measure $\big[\mathcal{D}u(h)\mathcal{D}v_\rho^{*}\,\mathcal{D} v_\sigma\big]$ in\,\eqref{meas} can be written as
\begin{equation}\label{meas2}
	\big[\mathcal{D}u(h)\mathcal{D} v_\rho^{*}\,\mathcal{D} v_\sigma\big]= \mathcal{A}\prod_x\Big[\Big(\prod_{\alpha\leq\,\beta}\dd{\widehat h_{\alpha\beta}(x)}\Big)\Big(\prod_{\rho}\dd{\widehat v_\rho^{\,*}(x)}\Big)\Big(\prod_{\sigma}\dd{\widehat v_\sigma(x)}\Big)\Big]\,,
\end{equation}
where $a$-independent terms such as $\prod_x \widetilde g^{\,00}(x)\,\left(\widetilde g(x)\right)^{-1}$ are included in $\mathcal{A}$. 

Since equations \eqref{S2+Sgf} and\,\eqref{ghostaction2} contain the {\it dimensionless} operators $-\nabla_\rho\nabla^{\rho}$ and $-\nabla_\rho$, to calculate the path integral in\,\eqref{effacgrav1} we consider the bases for symmetric tensors and for vectors constructed with the eigenfunctions of the dimensionless Laplace-Beltrami operator $-\widetilde\square^{(s)}$ defined as (see Appendix A for details)
\begin{equation}\label{LB1}
-\widetilde\square^{(s)}\equiv-a^2\,\square^{(s)}_{\,a}\,, 
\end{equation}
where $-\square^{(s)}_a$ are the spin-$s$ Laplace-Beltrami operators for a sphere of radius $a$, with $s=0,1,2$. The dimensionless eigenvalues $\lambda_n^{(s)}$ of $-\widetilde\square^{(s)}$ and the corresponding degeneracies $D_n^{(s)}$ are 
\begin{equation}
	\lambda_n^{(s)}=n^2+3n-s\qquad ;\qquad D_n^{(s)}=\frac{2s+1}{3}\left(n+\frac{3}{2}\right)^3-\frac{(2s+1)^3}{12}\left(n+\frac{3}{2}\right)\,,
	\label{eigenvalues}
\end{equation}
where $n=s,s+1,\dots$\,. Expanding $\widehat h_{\mu\nu}$, $\widehat v^{\,*}_\rho$ and $\widehat v_\sigma$ in terms of the bases mentioned above, $\delta S^{1l}_{\rm grav}$ turns out to be 
\begin{align}\label{result}
	\delta S^{1l}_{\rm grav}&=-\frac12\log\frac{\det_1 [-\widetilde\square^{(1)}-3]\det_2 [-\widetilde\square^{(0)}-6]}{\det_0 [-\widetilde\square^{(2)}-2a^2\Lambda_{\rm cc}+8]\det_2 [-\widetilde\square^{(0)}-2a^2\Lambda_{\rm cc}]}+\frac12\log(2a^2\Lambda_{\rm cc})+\mathcal{B}\,,
\end{align}
where $\mathcal{B}$ is an $a$-independent term, and the index $i$ in $\det_i$ indicates that the product of eigenvalues starts from $\lambda^{(s)}_{s+i}\,$. The calculations that lead to\,\eqref{result} are presented in Appendix A, and closely follow \cite{TaylorVeneziano}.

Eq.\,\eqref{result} is an important outcome of our calculation, and we pause for a moment to comment on it. The term {\small $\frac12\log(2a^2\Lambda_{\rm cc})$} comes from the integration over one of the modes of {\small $\widehat h_{\mu\nu}$} (see Appendix A), $\mathcal B$ (as already said) is an $a$-independent term. Both terms are irrelevant for our scopes. The only relevant term in\,\eqref{result} is the first one. Its peculiarity is that it contains only {\it dimensionless determinants}.
This is a fundamental result of our analysis and is due to the fact that we have appropriately considered the correct measure\,\eqref{meas}. In particular, we stress the importance of the term \,{\small $g^{(a)\,00}(x)\,\left(g^{(a)}(x)\right)^{-1}$} whose presence is sometimes overlooked. As Eq.\,\eqref{measterms} shows, in fact, it provides the factor $a^{-10}$ that compensates the factor $a^{10}$ coming from {\small $	\prod_{\alpha\leq\,\beta}\dd{h_{\alpha\beta}(x)}$} in Eq.\,\eqref{prodh}. It is also worth to stress that, when written in terms of \,{\small $\widehat h_{\mu\nu}$} and $\widehat v_{\mu}$, $\delta S^{(2)}$ contains only dimensionless quantum fluctuation operators, see\,\eqref{S2+Sgf} and\,\eqref{ghostaction2}. In this respect, we note that in typical calculations of $\delta S^{1l}_{\rm grav}$ the arguments in ${\rm det}_i$ are dimensionful, and an arbitrary scale $\mu$ (supposed to be harmless) is introduced to make them dimensionless. In our result\,\eqref{result}, the determinants turned out to be {\it automatically}  dimensionless\footnote{Note that the radius $a$ and the cosmological constant $\Lambda_{\rm cc}$ appear only in the dimensionless combination $a^2\Lambda_{\rm cc}$.}, and no such arbitrary scale is ever needed. 

We now move to the calculation of the right hand side of\,\eqref{result} (keeping as explained above only the first term). We will follow two different strategies: (i) we perform the calculation of the determinants performing the product of a finite number $N$ of eigenvalues; (ii) we repeat the calculation using proper-time regularization. Anticipating on the results, from both calculations we will see that the quartically and quadratically divergent contributions to the vacuum energy are absent. We will also explain why in the literature these terms are usually found. 

Let us begin with the direct calculation of the determinants in terms of product of eigenvalues. From\,\eqref{result} we have
\begin{align}
	\delta S^{1l}_{\rm grav}=\frac12\sum_{n=2}^{N-2} &\Bigl[D_n^{(2)}\log\left(\lambda_n^{(2)}-2a^2\Lambda_{\rm cc}+8\right)+D_n^{(0)}\log\left(\lambda_n^{(0)}-2a^2\Lambda_{\rm cc}\right)\nonumber\\
	&-D_n^{(1)}\log\left(\lambda_n^{(1)}-3\right)-D_n^{(0)}\log\left(\lambda_n^{(0)}-6\right)\Bigr]+\frac12\log(2a^2\Lambda_{\rm cc})+\mathcal{B}\,,
	\label{calculation1}
\end{align}
where the UV cutoff is introduced in terms of a numerical cut $N$ ($N\gg1$) on the number of eigenvalues. The choice $N-2$ (rather than $N$) in the upper limit of the sum is just a matter of convenience and simplifies the expression of\,  $\delta S^{1l}_{\rm grav}$.

Before going on with the calculation, it is worth to stress that a similar numerical cut was introduced in\,\cite{Becker:2020mjl,Becker:2021pwo,Ferrero:2024yvw}. In particular, in \cite{Becker:2021pwo} the same pure gravity case considered in the present section is studied, while \cite{Becker:2020mjl} and \cite{Ferrero:2024yvw} are devoted to the case of a scalar theory in a gravitational background, a setup that we will consider in section 4.  There is however  a crucial difference between the calculation performed in \cite{Becker:2021pwo} and our calculation. As shown by Fradkin and Vilkovisky\cite{Fradkin:1973wke},  the measure to be used to calculate $\delta S^{1l}_{\rm grav}$  in \eqref{effacgrav1} is given by \eqref{meas}, and  results from the integration over the conjugate momenta of the fields  (that appear in the path integral of the original Hamiltonian formulation of the theory). The measure considered in \cite{Becker:2021pwo}, instead, cannot be derived from such a functional integration, and therefore it is not suitable to calculate the effective action. It is not difficult to see that this incorrect choice of the measure leads the authors to conclusions that ought to be reconsidered. Similar considerations\footnote{As already mentioned,  in \cite{Becker:2020mjl} and \cite{Ferrero:2024yvw} the case of a scalar field in a gravitational background is studied. Similarly to \cite{Becker:2021pwo}, in \cite{Becker:2020mjl} the measure is not the one derived from the original path integral in hamiltonian formalism, while in \cite{Ferrero:2024yvw} the metric-dependent terms of the measure (see for instance \eqref{meas}) are not taken into account.} apply to \cite{Becker:2020mjl} and \cite{Ferrero:2024yvw}.

Let us go back to the explicit calculation of $\delta S^{1l}_{\rm grav}$. Inserting\,\eqref{eigenvalues} in the right hand side of\,\eqref{calculation1} and using the identity {\small $\log\left(x/y\right)=-\int_{0}^{+\infty}\dd{u}\left[\left(x+u\right)^{-1}-\left(y+u\right)^{-1}\right]$}, the sum can be put in closed form, though the expression is quite involved. For our purposes, it is sufficient to consider the expansion for $N\gg 1$, that gives
\begin{align}
	\delta S^{1l}_{\rm grav}=\,&-\left(\Lambda_{\rm cc}^2 \log N^2\right)a^4+ \Lambda_{\rm cc}\left(-N^2+8\log N^2\right)a^2\nonumber\\
	&+\frac{N^4}{24}\left(-1+2\log N^2\right)+\frac{N^2}{36}\left(203-75\log N^2\right)-\frac{779}{90}\log N^2+\mathcal{B}\nonumber\\
	&+\frac12\log(2a^2\Lambda_{\rm cc})+\mathcal F(a^2\Lambda_{\rm cc})+\mathcal{O}\left(N^{-2}\right)\,,
	\label{oneloopres}
\end{align}
where $\mathcal F(a^2\Lambda_{\rm cc})$ contains only UV-finite terms (no dependence on $N$), and its explicit expression is given in Appendix B.
As already said (see comments below\,\eqref{ghostaction}), the one-loop corrections $\delta\left(\frac{\Lambda_{\rm cc}}{G}\right)$ and $\delta\left(\frac{1}{G}\right)$ to $\frac{\Lambda_{\rm cc}}{G}$ and $\frac{1}{G}$ are to be identified with the coefficients of $a^4$ and $a^2$ in\,\eqref{oneloopres} respectively. The cosmological and Newton constant at one-loop turn then out to be
\begin{align}
	\Lambda^{1l}_{\rm cc}=&\frac{\Lambda_{\rm cc}\left(1-\frac{3G\Lambda_{\rm cc}}{\pi} \log N^2\right)}{1+\frac{G\Lambda_{\rm cc}}{2\pi}\left(N^2-8\log N^2\right)}\label{CC}\\
	G^{1l}=&\frac{G}{1+\frac{G\Lambda_{\rm cc}}{2\pi}\left(N^2-8\log N^2\right)}\,.
	\label{NC1}
\end{align}
Since $\Lambda_{\rm cc}^{1l}$ and $G^{1l}$ are the renormalized values of the cosmological and Newton constant, we must have $\Lambda_{\rm cc}^{1l}>0$ and $G^{1l}>0$. A simple inspection of\,\eqref{CC} and\,\eqref{NC1} shows that only positive values of the bare parameters $\Lambda_{\rm cc}$ and $G$ are then admitted. 

We now discuss the relation between the numerical cut $N$ and the physical UV cutoff $\Lambda_{\rm cut}$. From \,\eqref{bareaction}, the classical (de Sitter) solution
\begin{equation}\label{desitterbare}
	a_{_{\rm dS}}=\sqrt{\frac{3}{\Lambda_{\rm cc}}}
\end{equation}
is obtained and, since $a_{_{\rm dS}}$ is the (tree-level approximation to the) size of the universe, the connection between $N$ and $\Lambda_{\rm cut}$ (that might be for instance $\mpl$) is given by
\begin{equation}\label{MpN}
	\Lambda_{\rm cut}=N\sqrt{\frac{\Lambda_{\rm cc}}{3}}\,.
\end{equation}
Since the physical cutoff $\Lambda_{\rm cut}$ is a fixed scale, from\,\eqref{MpN} we see that for increasing values of $a_{_{\rm dS}}$ (i.e. decreasing values of $\Lambda_{\rm cc}$) $N$ grows linearly with $a_{_{\rm dS}}$. This is clearly as expected: to keep the minimal distance resolution ($\sim1/\Lambda_{\rm cut}$) unchanged for increasing values of $a_{_{\rm dS}}$, a finer angular resolution is needed. This requires the inclusion of a larger number of eigenmodes in the decomposition of the fluctuation fields. In other words, for a given value of the physical cutoff $\Lambda_{\rm cut}$ the number $N$ of eigenvalues to be included in the sum\,\eqref{calculation1} is fixed by the highest eigenvalue $\sim N^2/a_{_{\rm dS}}^2$ of the Laplace-Beltrami operator for the sphere of radius $a_{_{\rm dS}}$. 

In this respect, it is worth to recall that for a sphere of generic radius $a$ the dimensionful eigenvalues {\small $\widehat\lambda_n^{\,(s)}$} of the Laplace-Beltrami operator $-\square_{\,a}^{(s)}$ are $\widehat\lambda_n^{\,(s)}\equiv\frac{\lambda_n^{(s)}}{a^2}\sim\frac{n^2}{a^2}$ (see\,\eqref{eigenvalues} where $\lambda_n^{(s)}$ are the eigenvalues of the dimensionless operators $-\widetilde\square^{(s)}$). It might seem natural at first to introduce the physical cutoff $\Lambda_{\rm cut}$ through the requirement $\widehat\lambda_n^{\,(s)}\leq\Lambda_{\rm cut}^2$, rather than through\,\eqref{MpN}. 
However, as the eigenvalues $\widehat \lambda^{\,(s)}_{n}$, that {\it are not} the eigenvalues of the operators appearing in the determinants in\,\eqref{result}, are proportional to $a^{-2}$, such a requirement would introduce  in $N$ a spurious dependence on $a$, i.e.\,\,a spurious dependence on the classical background metric $g^{(a)}_{\mu\nu}$. In other words, the identification { $\Lambda_{\rm cut}^2\sim \widehat \lambda^{\,(s)}_{\rm max}\sim N^2/a^2$} would give rise to spurious additional terms $\sim a^4$ and $\sim a^2$ in $\delta S^{1l}_{\rm grav}$. Consequently, the results for the one-loop corrections to $\frac{\Lambda_{\rm cc}}{G}$ and $\frac{1}{G}$ would be altered. In particular, as we will show below, this artificially generates the quartically and quadratically divergent contributions to the vacuum energy $\rho_{\rm vac}=\frac{\Lambda_{\rm cc}}{8\pi G}$ usually found in the literature.

Finally, inserting\,\eqref{MpN} in\,\eqref{oneloopres}, for $\delta S^{1l}_{\rm grav}$ we have
\begin{align}
	\delta S^{1l}_{\rm grav}=\,&-\left(\Lambda_{\rm cc}^2 \log\frac{3\Lambda_{\rm cut}^2}{\Lambda_{\rm cc}}\right)a^4+ \left(-3\Lambda_{\rm cut}^2+8\Lambda_{\rm cc}\log\frac{3\Lambda_{\rm cut}^2}{\Lambda_{\rm cc}}\right)a^2\nonumber\\
	&+\frac{3\Lambda_{\rm cut}^4}{8\Lambda_{\rm cc}^2}\left(-1+2\log\frac{3\Lambda_{\rm cut}^2}{\Lambda_{\rm cc}}\right)+\frac{\Lambda_{\rm cut}^2}{12\Lambda_{\rm cc}}\left(203-75\log\frac{3\Lambda_{\rm cut}^2}{\Lambda_{\rm cc}}\right)-\frac{779}{90}\log\frac{3\Lambda_{\rm cut}^2}{\Lambda_{\rm cc}}+\mathcal B\,\,\,\,\,\,\,\nonumber\\
	&+\frac12\log(2a^2\Lambda_{\rm cc})+\mathcal F(a^2\Lambda_{\rm cc})+\mathcal{O}\left(\Lambda_{\rm cut}^{-2}\right)\,.
	\label{oneloopressub}
\end{align}
Eq.\,\eqref{oneloopressub} (and equivalently\,\eqref{oneloopres}) is one of the most important findings of the present work. We will comment on the consequences of this result in the next section. Before doing that, it is useful to proceed with the evaluation of $\delta S^{1l}_{\rm grav}$ following the other strategy mentioned above, namely proper-time regularization. We will then conveniently discuss both results together. 

Since the operators $(-\widetilde\square^{(s)}-\alpha)$ in\,\eqref{result} (with $\alpha=3,\,6,\,2a^2\Lambda_{\rm cc}-8,\,2a^2\Lambda_{\rm cc}$) are dimensionless, to regularize the determinants ${\rm det}_i(-\widetilde\square^{(s)}-\alpha)$ we introduce the dimensionless proper-time  $\tau$, where the lower cut in the integral is a number $N\gg1$
\begin{equation}\label{propertime}
	{\rm det}_{i}(-\widetilde\square^{(s)}-\alpha)=e^{-\int_{1/N^2}^{+\infty}\frac{\dd{\tau}}{\tau}\,{\rm K}_i^{(s)}(\tau)}\,.
\end{equation}
The kernel ${\rm K}_i^{(s)}(\tau)$ is ($\lambda^{(s)}_n$ and $D^{(s)}_n$ are the eigenvalues and the degeneracies in\,\eqref{eigenvalues})
\begin{equation}\label{kernel}
	{\rm K}_i^{(s)}(\tau)= \sum_{n=s+i}^{+\infty}D^{(s)}_n\, e^{-\tau\left(\lambda^{(s)}_n-\alpha\right)}\,_{_\text{\normalsize .}}
\end{equation}
To calculate the determinants, we insert\,\eqref{kernel} in\,\eqref{propertime}, perform the integration over $\tau$, and then sum over $n$ with the help of the EML formula. For the reader's convenience, we report here this formula
\begin{align}
	\sum_{n=n_i}^{n_f}f(n)=\int_{n_i}^{n_f}\dd[]x\,f(x)+\frac{f(n_f)+f(n_i)}{2}+\sum_{k=1}^{p}\frac{B_{2k}}{(2k)!}\left(f^{(2k-1)}(n_f)-f^{(2k-1)}(n_i)\right)+R_{2p}\,,
	\label{EML}
\end{align}
where $p$ is an integer, $B_m$ are Bernoulli numbers and the rest $R_{2p}$ is given by
\begin{align}
	\small
	R_{2p}=\sum_{k=p+1}^{\infty}\frac{B_{2k}}{(2k)!}\left(f^{(2k-1)}(n_f)-f^{(2k-1)}(n_i)\right)=\frac{(-1)^{2p+1}}{(2p)!}\int_{n_i}^{n_f}\dd[]x\,f^{(2p)}(x)B_{2p}(x-[x]),
	\label{rest}
\end{align}
with $B_n(x)$ the Bernoulli polynomials, $[x]$ the integer part of $x$, and $f^{(i)}$ the $i$-th derivative of $f$ with respect to its
argument. 

Expanding for $N\gg 1$, we finally get
\begin{align}
	\delta S^{1l}_{\rm grav}=\, &-\left(\Lambda_{\rm cc} ^2\log N^2\right)a^4+\Lambda_{\rm cc}\left(-N^2+8\log N^2\right)a^2\nonumber\\
	&-\frac{N^4}{12}+\frac{17 }{3}N^2-\frac{1859}{90}\log N^2+\mathcal{B}\nonumber\\
	&+\frac12\log(2a^2\Lambda_{\rm cc})+\mathcal{G}(a^2\Lambda_{\rm cc})+\mathcal{O}\left(N^{-2}\right)\,,
	\label{oneloopresHK}
\end{align}
where $\mathcal G(a^2\Lambda_{\rm cc})$ contains only UV-finite terms (no dependence on $N$), and is similar to the term $\mathcal F(a^2\Lambda_{\rm cc})$ in\,\eqref{oneloopres}. As for the previous calculation, the connection between the numerical cut $N$ and the physical cutoff $\Lambda_{\rm cut}$ is given by the relation (see\,\eqref{MpN} and comments below)
\begin{equation}\label{Lambdapt}
	\Lambda_{\rm cut}\equiv\frac{N}{a_{_{\rm dS}}}=\sqrt{\frac{\Lambda_{\rm cc}}{3}}\,N\,,
\end{equation}
so that\,\eqref{oneloopresHK} is written as
\begin{align}
	\delta S^{1l}_{\rm grav}=\,&-\left(\Lambda_{\rm cc}^2 \log\frac{3\Lambda_{\rm cut}^2}{\Lambda_{\rm cc}}\right)a^4+ \left(-3\Lambda_{\rm cut}^2+8\Lambda_{\rm cc}\log\frac{3\Lambda_{\rm cut}^2}{\Lambda_{\rm cc}}\right)a^2\nonumber\\
	&-\frac{3\Lambda_{\rm cut}^4}{4\Lambda_{\rm cc}^2}+\frac{17\Lambda_{\rm cut}^2}{\Lambda_{\rm cc}}-\frac{1859}{90}\log\frac{3\Lambda_{\rm cut}^2}{\Lambda_{\rm cc}}+\mathcal B\,\,\,\,\,\,\,\nonumber\\
	&+\frac12\log(2a^2\Lambda_{\rm cc})+\mathcal G(a^2\Lambda_{\rm cc})+\mathcal{O}\left(\Lambda_{\rm cut}^{-2}\right)\,.
	\label{onelooprespt}
\end{align}
In summary, for $\delta S^{1l}_{\rm grav}$ we have found\,\eqref{onelooprespt} (or equivalently\,\eqref{oneloopresHK}) using the proper-time method, and correspondingly\,\eqref{oneloopressub} (or\,\eqref{oneloopres}) resorting to the direct product of eigenvalues. 

In the next section, we will focus on the most important consequences of Eqs.\,\eqref{oneloopressub} and\,\eqref{onelooprespt}, considering the corrections to $\frac{\Lambda_{\rm cc}}{G}$ and $\frac{1}{G}$ that result from these equations. We will also explain the reason why in previous literature quartically and quadratically divergent contributions to the vacuum energy are typically found. 

\section{One-loop corrections to $\frac{\Lambda_{\rm cc}}{G}$ and $\frac{1}{G}$}
\label{oneloop}

As already said, the coefficients of $a^4$ and $a^2$ in $\delta S^{1l}_{\rm grav}$ provide the corrections to $\frac{\Lambda_{\rm cc}}{G}$ and $\frac{1}{G}$ respectively. In the previous section, we calculated $\delta S^{1l}_{\rm grav}$ resorting first to the direct product of eigenvalues, that led to\,\eqref{oneloopressub}, and then to the proper-time method, that resulted in\,\eqref{onelooprespt}. Though they were derived using two different methods, from both\,\eqref{oneloopressub} and\,\eqref{onelooprespt} we find that up to one-loop order
\begin{align}
	\frac{\Lambda^{1l}_{\rm cc}}{G^{1l}}=&\frac{\Lambda_{\rm cc}}{G}\left(1-\frac{3G\Lambda_{\rm cc}}{\pi} \log\frac{3 \Lambda_{\rm cut}^2}{\Lambda_{\rm cc}}\right)\label{VE}\\
	\frac{1}{G^{1l}}=&\frac{1}{G}\left[1+\frac{G}{2\pi}\left(3 \Lambda_{\rm cut}^2-8\Lambda_{\rm cc}\log \frac{3 \Lambda_{\rm cut}^2}{\Lambda_{\rm cc}}\right)\right]\,.
	\label{NC}
\end{align}

We now comment on\,\eqref{VE} and\,\eqref{NC}, starting from the latter. First of all we observe that, since $\Lambda_{\rm cut}\sim\mpl$, taking for the Newton constant the \vv natural value" $G\sim\mpl^{-2}$, from\,\eqref{NC} we see that the dressing of $G$ does not spoil this natural relation, 
\begin{equation}\label{Gren}
	G^{1l}\sim G\sim\frac{1}{\mpl^2}\,.
\end{equation}
In other words, there is no naturalness problem in connection with the renormalization of the Newton constant.

Moving to\,\eqref{VE}, we immediately see that it contains a surprising result: quantum fluctuations dress the vacuum energy $\rho_{\rm vac}=\frac{\Lambda_{\rm cc}}{8\pi G}$ \,{\it only}\, with logarithmic corrections\footnote{As we show in the next section, this holds true also when matter is included.}. The quantum correction to $\rho_{\rm vac}$ \,goes like \,$\log\mpl$ (again we take $\Lambda_{\rm cut}\sim\mpl$) \,rather than $\mpl^4$, the latter being the typically acknowledge UV-sensitivity. This strong power-like dependence on the UV scale requires the bare value $\frac{\Lambda_{\rm cc}}{8\pi G}$ of the vacuum energy to be $\sim\mpl^4$, with a coefficient that must be enormously fine-tuned for it to cancel (quite exactly) the one-loop generated $\mpl^4$ correction. On the contrary, with Eq.\,\eqref{VE} the bare cosmological constant $\Lambda_{\rm cc}$ does not need to be $\sim\mpl^2$, but might well be $\Lambda_{\rm cc}\ll\mpl^2$. Under this latter condition, from\,\eqref{VE} and\,\eqref{Gren} we have
\begin{equation}\label{Lambdaren}
	\Lambda^{1l}_{\rm cc}\sim\Lambda_{\rm cc}\,.
\end{equation}
As a consequence, in pure gravity there is {\it no naturalness problem} for the cosmological constant. Note also that\,\eqref{Lambdaren} implies that bare and dressed de Sitter radius practically coincide.

It is important at this point to understand why, when the calculation is performed with techniques (proper-time regularization within the heat kernel expansion) similar to those we used in the previous section, quartically and quadratically divergent contributions to the vacuum energy are typically found (obviously dimensional regularization is excluded from these considerations since the method itself is constructed ad hoc not to display them).
To investigate on this point, we consider\,\eqref{oneloopresHK} (equivalently we could consider\,\eqref{oneloopres}), and for the sake of the present discussion we temporarily realize the connection between $N$ and $\Lambda_{\rm cut}$ through the relation 
\begin{equation}\label{incorrectLambdapt}
\Lambda_{\rm cut}=\frac{N}{a}\,,
\end{equation}
rather than via $\Lambda_{\rm cut}=N/a_{_{\rm dS}}$ (Eq.\,\eqref{Lambdapt}). As explained above, Eq.\,\eqref{incorrectLambdapt} would correspond to the (improper) identification of the dimensionful cutoff $\Lambda_{\rm cut}$ with the maximal eigenvalue {\small $\widehat \lambda_{\rm max}^{(s)}$} of the dimensionful Laplacian $-\square^{(s)}_{\,a}$ (see comments below\,\eqref{MpN} and above\,\eqref{Lambdapt}). Inserting\,\eqref{incorrectLambdapt} in\,\eqref{oneloopresHK}, for $\delta S^{1l}_{\rm grav}$ we would obtain (we neglect the terms starting from $\mathcal B$ as they are inessential for the present discussion)
\begin{align}
	\delta S^{1l}_{\rm grav}=\, &-\left[\Lambda_{\rm cc} ^2\log\left(a^2\Lambda_{\rm cut}^2\right)\right]a^4+\Lambda_{\rm cc}\left[-\Lambda_{\rm cut}^2\,a^2+8\log\left(a^2\Lambda_{\rm cut}^2\right)\right]a^2\nonumber\\
	&-\frac{\Lambda_{\rm cut}^4}{12}\,a^4+\frac{17}{3}\Lambda_{\rm cut}^2\,a^2-\frac{1859}{90}\log\left(\Lambda_{\rm cut}^2\,a^2\right)\,,
	\label{incorrectident}
\end{align}
which is trivially rewritten as
\begin{align}
	\delta S^{1l}_{\rm grav}=\, &-\left[\frac{\Lambda_{\rm cut}^4}{12}+\Lambda_{\rm cc}\Lambda_{\rm cut}^2+\Lambda_{\rm cc} ^2\log\left(a^2\Lambda_{\rm cut}^2\right)\right]a^4+\left[\frac{17}{3}\Lambda_{\rm cut}^2+8\Lambda_{\rm cc}\log\left(a^2\Lambda_{\rm cut}^2\right)\right]a^2\nonumber\\
	&-\frac{1859}{90}\log\left(a^2\Lambda_{\rm cut}^2\right)\,.
	\label{incorrectident2}
\end{align}
Eq.\,\eqref{incorrectident2} is nothing but the well-known result found in the literature when the calculation is performed resorting to proper-time regularization within the heat kernel expansion. Quartically and quadratically divergent corrections to $\frac{\Lambda_{\rm cc}}{G}$ are found.

The comparison between Eq.\,\eqref{incorrectident} (that we have written only for the sake of the present discussion using Eq.\,\eqref{incorrectLambdapt} for the physical cutoff $\Lambda_{\rm cut}$) and the original result\,\eqref{oneloopresHK} for $\delta S^{1l}_{\rm grav}$ allows to understand how these spurious divergences are generated. This point is crucial to our analysis, and it is worth to examine these terms one after the other. Taking the term $-\Lambda_{\rm cc}\,N^2\,a^2$ of\,\,\eqref{oneloopresHK}, and replacing in it $N^2$ according to\,\eqref{incorrectLambdapt}, the quadratically divergent term $-\Lambda_{\rm cc}\,\Lambda_{\rm cut}^2\,a^4$ of\,\,\eqref{incorrectident2} arises. This is an example of how, when the UV cutoff $\Lambda_{\rm cut}$ is improperly identified through\,\eqref{incorrectLambdapt}, a term that is originally proportional to $a^2$ (artificially) becomes an $a^4$ term. Consequently, a term that originally renormalizes $\frac{1}{G}$ becomes a quadratic divergence that renormalizes $\frac{\Lambda_{\rm cc}}{G}$. Similarly, $-\frac{N^4}{12}$ in the second line of\,\eqref{oneloopresHK} becomes {\footnotesize $-\frac{\Lambda_{\rm cut}^4}{12}\,a^4$}, thus (artificially) giving rise to the (in)famous quartically divergent contribution to the vacuum energy {\small $\frac{\Lambda_{\rm cc}}{8\pi G}$}. Finally, {\footnotesize $\frac{17}{3}N^2$} in the second line of\,\eqref{oneloopresHK} becomes {\footnotesize $\frac{17}{3}\Lambda_{\rm cut}^2\,a^2$}, i.e.\,\,a quadratically divergent contribution to $\frac{1}{G}$. Concerning this latter term, a simple inspection of\,\eqref{onelooprespt} and\,\eqref{incorrectident2} shows that\,\eqref{incorrectLambdapt} generates in the one-loop correction to $\frac{1}{G}$ a quadratic divergence that is opposite in sign with respect to the original one (Eq.\,\eqref{onelooprespt}).

The importance of the above results and considerations can hardly be underestimated. What we have just seen is that implementing the cut in the fluctuation determinants taking as physical cutoff the maximal eigenvalues $\widehat \lambda_{\rm max}^{(s)}$ (see\,\eqref{incorrectLambdapt}) introduces in $\delta S^{1l}_{\rm grav}$ a spurious dependence on the background metric $g_{\mu\nu}^{\text{\tiny $(a)$}}$. As stressed above, the connection between the numerical cut $N$ and the physical cutoff $\Lambda_{\rm cut}$ must rather be realised through\,\eqref{Lambdapt}, i.e. through the de Sitter radius $a_{_{\rm dS}}$. 

Before ending this section, we comment on the terms in the second and third line of Eq.\,\eqref{oneloopresHK}. Starting with the third line, we observe that $\frac12\log(2a^2\Lambda_{\rm cc})$ and $\mathcal G(a^2\Lambda_{\rm cc})$ are $\mathcal O(1)$, negligible contributions to $\delta S_{\rm grav}^{1l}$ (the same holds true for the similar terms $\frac12\log(2a^2\Lambda_{\rm cc})$ and $\mathcal F(a^2\Lambda_{\rm cc})$ in the third line of\,\eqref{oneloopres}). Concerning the terms in the second line, in our analysis we interpret them as constants, though the high symmetry of the spherical background prevents from a clear distinction between constants and $R^2$ terms. A less symmetric background should help in clarifying this issue.

In the present and previous section we considered pure gravity, ignoring the matter term $S_{\rm mat}$ in the action. A full treatment of the quantum matter-gravity system is beyond the scopes of the present work, and we leave it for future studies \cite{noi}. Still, we can have an indication of the role that matter plays on the renormalization of $\frac{\Lambda_{\rm cc}}{G}$ and $\frac{1}{G}$ considering the simpler cases of free scalar and fermion fields on a gravitational background. The next section is devoted to this analysis.

\section{Matter contribution}
\label{matter}
\subsection{Free scalar field}
\label{scalar}

Let us consider the free theory\footnote{Work on the interacting theory is in progress \cite{interacting}.} of a real scalar field $\phi$ of mass $m_\phi$ defined on the spherical gravitational background $g^{(a)}_{\mu\nu}$,
\begin{equation}
	S_{\rm mat}^{(\phi)}=\int\dd[4]x\,\sqrt{g^{(a)}}\left[\frac12\,g^{(a)\,\mu\nu} \partial_\mu\phi\,\partial_\nu\phi+\frac12\,m_\phi^2\,\phi^2\right]\,.
	\label{scalaraction2}
\end{equation}
Adding $S^{(\phi)}_{\rm mat}$ to the classical gravitational action $S^{(a)}_{\rm grav}$ in\,\eqref{Sb}, the total action is
\begin{equation}
	S=S^{(a)}_{\rm grav}+S_{\rm mat}^{(\phi)}=\frac{\pi\Lambda_{\rm cc}}{3G}a^4-\frac{2\pi}{G}a^2+\int\dd[4]x\,\sqrt{g^{(a)}}\left[\frac12\,g^{(a)\,\mu\nu} \partial_\mu\phi\,\partial_\nu\phi+\frac12\,m_\phi^2\,\phi^2\right]\,.
	\label{totaction}
\end{equation}
Integrating out the field $\phi$, for the effective action we get
\begin{equation}
	S_{\rm grav}^{\rm eff}=\frac{\pi\Lambda_{\rm cc}}{3G}a^4-\frac{2\pi}{G}a^2+\delta S_{\rm grav}\,,
\end{equation}
with $\delta S_{\rm grav}$ given by
\begin{align}\label{effac1}
	e^{-\delta S_{\rm grav}}=\int\prod_x\left[\left(g^{(a)\,00}(x)\right)^{\frac12}\left(g^{(a)}(x)\right)^{\frac14}\dd{\phi\text{\footnotesize $(x)$}}\right] e^{-\text{\small $\int$}\dd[4]x\, \sqrt{g^{(a)}}\,\left[-\frac12 \phi\,\,\square^{(0)}_{\,a}\,\phi+\frac12m_\phi^2\,\phi^2\right]}\,.
\end{align}

As for pure gravity, the factors {\small $\left(g^{(a)\,00}(x)\right)^{\frac12}\left(g^{(a)}(x)\right)^{\frac14}$} in the measure come from the integration over the conjugate momenta in the original path integral (hamiltonian formalism)\,\cite{Fradkin:1973wke,Unz:1985wq}. Similarly to that case, their presence is crucial to obtain the correct result for $\delta S_{\rm grav}$. Since from\,\eqref{metric1}
\begin{equation}
	\left(g^{(a)\,00}(x)\right)^{\frac12}\left(g^{(a)}(x)\right)^{\frac14}=a\left(\widetilde g^{\,00}(x)\right)^{\frac12}\left(\widetilde g(x)\right)^{\frac14}\,,
\end{equation}
Eq.\,\eqref{effac1} can be rewritten in terms of dimensionless quantities as
\begin{align}\label{effac2}
	e^{-\delta S_{\rm grav}}=\mathcal{A}\int\prod_{x}\left[\dd{\widehat\phi\text{\footnotesize $(x)$}}\right] e^{-\text{\small $\int$}\dd[4]x\,\sqrt{\,\widetilde g\,\,}\left[-\frac12\,\widehat\phi\,\,\widetilde\square^{(0)}\,\widehat\phi\,+\,\frac12a^2\,m_\phi^2\,\widehat\phi^{\,2}\right]}\,,
\end{align}
where $-\widetilde\square^{(0)}$ is the dimensionless spin-$0$ Laplacian (see\,\eqref{LB1}), $\widehat \phi$\, the dimensionless field defined as $\widehat\phi\equiv a\phi$, and $\mathcal A$ an $a$-independent factor. 

Expanding now \,$\widehat\phi\text{\footnotesize $(x)$}$\, in terms of the eigenfunctions\footnote{The $\phi_n^{(i)}$ are normalized as $\int\dd[4]{x}\sqrt{\,\widetilde g\,\,}\,\,\phi_n^{(i)}(x)\,\phi_m^{(j)}(x)=\delta^{ij}\delta_{nm}$.} $\phi_n^{(i)}(x)$ ($i$ degeneracy index and $n=0, 1, \dots$) of the operator $-\widetilde\square^{(0)}$ (the eigenvalues of $-\widetilde\square^{(0)}$, with the corresponding degeneracies, are given in\,\eqref{eigenvalues}), i.e.\,writing \,{\small $\widehat\phi=\sum_{n,i}c_n^{(i)}\phi_n^{(i)}$}\,,
from\,\eqref{effac2} we have
\begin{align}
	e^{-\delta S_{\rm grav}}=\mathcal{A}\int\prod_{n,i}\dd{c_n^{(i)}} e^{-\frac12\sum_{n,i}\left[c_n^{(i)}\right]^2\left(\lambda_n^{(0)}+a^2m_\phi^2\right)}\,,
	\label{oneloopeffac2}
\end{align}
and then ($\mathcal{C}$ is an $a$-independent term)
\begin{equation}\label{scalarres}
	S_{\rm grav}^{\rm eff}=\frac{\pi\Lambda_{\rm cc}}{3G}a^4-\frac{2\pi}{G}a^2+\frac12\log\left[{\rm det}\left(-\widetilde\square^{(0)}+a^2 m_\phi^2\right)\right]+\,\mathcal{C}\,.
\end{equation}
It is worth to note that, had we missed the factors {\small $\left(g^{(a)\,00}(x)\right)^{\frac12}\left(g^{(a)}(x)\right)^{\frac14}$} in the measure (see\,\eqref{effac1}), the $a$-dependence of the determinant in\,\eqref{scalarres} would have been altered. This would have made it dimensionful, thus requiring the introduction of an arbitrary scale $\mu$ to have a dimensionless argument in the logarithm (see comments in the paragraph below Eq.\,\eqref{calculation1}).

To calculate the determinant in\,\eqref{scalarres}, we use the proper-time method (see\,\eqref{propertime}). Being the operator $(-\widetilde\square^{(0)}+a^2 m_\phi^2)$ dimensionless, the determinant above is written in terms of a dimensionless proper-time $\tau$ with numerical cut $N\gg1$ 
\begin{equation}
	{\rm det}(-\widetilde\square^{(0)}+a^2 m_\phi^2)=e^{-\int_{1/N^2}^{+\infty}\frac{\dd{\tau}}{\tau}\,{\rm K}^{(0)}(\tau)}\,,\label{propertime*1}
\end{equation}
where the kernel ${\rm K}^{(0)}(\tau)$ is ($\lambda^{(0)}_n$ and $D^{(0)}_n$ are the eigenvalues and degeneracies reported in\,\eqref{eigenvalues})
\begin{equation}
	{\rm K}^{(0)}(\tau)= \sum_{n=0}^{+\infty}D^{(0)}_n\, e^{-\tau\left(\lambda^{(0)}_n+a^2m_\phi^2\right)}\,_{_\text{\normalsize .}}\label{propertime*2}
\end{equation}
Inserting\,\eqref{propertime*2} in\,\eqref{propertime*1}, performing the integration over $\tau$, summing over $n$ with the help of the EML formula, and finally expanding for $N\gg1$, we get
\begin{align}
	S_{\rm grav}^{\rm eff}=&\frac{\pi}{3}\left(\frac{\Lambda_{\rm cc}}{G}-\frac{m_\phi^4}{8\pi}\log N^2\right)a^4-2\pi\left[\frac{1}{G}- \frac{m_\phi^2}{24\pi}\left(N^2+2 \log N^2\right)\right]a^2\nonumber\\
	&-\frac{N^4}{24}-\frac{N^2}{6}-\frac{29}{180}\log N^2+\mathcal{C}+\text{finite}\,.
	\label{effaction2}
\end{align} 
The similarity of\,\,\eqref{effaction2} with\,\eqref{oneloopresHK} (and also with\,\eqref{oneloopres}) is evident. The numerical cut $N$ is related to the physical cutoff $\Lambda_{\rm cut}$ ($\sim\mpl$) through the relation\,\eqref{Lambdapt}, and as for the pure gravity case the vacuum energy $\rho_{\rm vac}=\frac{\Lambda_{\rm cc}}{8\pi G}$ receives only a (mild) logarithmically divergent correction,
\begin{equation}\label{QCve}
	\delta\left(\frac{\Lambda_{\rm cc}}{8\pi G}\right)=-\frac{m_\phi^4}{64\pi^2}\log \frac{3\Lambda_{\rm cut}^2}{\Lambda_{\rm cc}}\,.
\end{equation}

Now, had we proceeded with the (incorrect) identification of $\Lambda_{\rm cut}$ through the relation $\Lambda_{\rm cut}=\frac{N}{a}$ (see\,\eqref{MpN} and\,\eqref{incorrectLambdapt} and comments below them), spurious quartically and quadratically divergent terms would have appeared in the quantum correction to $\frac{\Lambda_{\rm cc}}{G}$ (see\,\eqref{incorrectident2} and comments below). For instance, with such an identification the term $-\frac{N^4}{24}$ would become $-\frac{\Lambda_{\rm cut}^4}{24}\,a^4$, that is a (spurious) quartically divergent correction to $\frac{\Lambda_{\rm cc}}{G}$; the term $\frac{N^2}{12}\,m^2\,a^2$ (that originally renormalizes $\frac{1}{G}$, see\,\eqref{effaction2}) would become $\frac{\Lambda_{\rm cut}^2}{12}\,m^2\,a^4$, thus generating an \vv artificial" quadratically divergent correction to $\frac{\Lambda_{\rm cc}}{G}$.

Eq.\,\eqref{QCve} is instructive also for another reason. If on the one hand it tells us that  in the quantum correction to the vacuum energy quartic and quadratic divergences do not appear, at the same time it shows that to fully solve the cosmological constant problem a physical mechanism that allows to get rid of the $\sim m_\phi^4$ contribution to $\frac{\Lambda_{\rm cc}}{G}$ has to be found. If $m_\phi\sim\mathcal{O}(\text{EW scale})$, and again we take $\Lambda_{\rm cut}\sim \mpl$, Eq.\,\eqref{QCve} shows that the cosmological constant problem is reduced from a $\sim 120$ orders of magnitude problem to a $\sim 50$ orders of magnitude one.

In the next section we continue our investigation on the matter contribution adding to $S^{(a)}_{\rm grav}$ the action $S_{\rm mat}^{(\psi)}$ of a free Dirac field on a spherical gravitational background.

\subsection{Free fermion field}
\label{fermion}

Let us consider the free theory of a massive Dirac field $\psi$ defined on the spherical gravitational background $g^{(a)}_{\mu\nu}$, whose action is ($\slashed{\nabla}$ is the Dirac operator on the sphere, $\slashed{\nabla}=\gamma^a e^{\mu}_a(\partial_\mu+\frac12\sigma^{bc}\omega_{bc\mu})$, where $\boldsymbol{e}_a\equiv e^{\mu}_a\partial_\mu$ and $\omega_{abc}\equiv e^{\mu}_a\,\omega_{bc\mu}$ are the vielbein and the spin connection on the sphere respectively)
\begin{equation}
	S_{\rm mat}^{(\psi)}=\int\dd[4]x\,\sqrt{g^{(a)}}\left[\frac12\left(\overline\psi(\slashed{\nabla}\psi)-\overline{(\slashed{\nabla}\psi)}\,\psi\right)+m_\psi\,\overline{\psi}\psi\right]\,.
	\label{spinoraction}
\end{equation}
Adding $S_{\rm mat}^{(\psi)}$ to $S^{(a)}_{\rm grav}$, the total action is
\begin{equation}
	S=S^{(a)}_{\rm grav}+S_{\rm mat}^{(\psi)}=\frac{\pi\Lambda_{\rm cc}}{3G}a^4-\frac{2\pi}{G}a^2+\int\dd[4]x\,\sqrt{g^{(a)}}\left[\frac12\left(\overline\psi(\slashed{\nabla}\psi)-\overline{(\slashed{\nabla}\psi)}\,\psi\right)+m_\psi\,\overline{\psi}\psi\right]\,.
	\label{spinortotaction}
\end{equation}

Integrating out the field $\psi$, the correction $\delta S_{\rm grav}$ to the classical Einstein-Hilbert action is given by \cite{Fradkin:1973wke,Donoghue:2020hoh,Honerkamp:1971xtx}
\begin{align}\label{effac1*}
	e^{-\delta S_{\rm grav}}=\int\prod_x\left[\left(g^{(a)}(x)\right)^{\frac38}\dd{\overline\psi(x)}\dd{\psi(x)}\right] e^{-\text{\small $\int$}\dd[4]x\,\sqrt{g^{(a)}}\,\left[\frac12\left(\overline\psi(\slashed{\nabla}\psi)-\overline{(\slashed{\nabla}\psi)}\,\psi\right)+m_\psi\,\overline{\psi}\psi\right]}\,.
\end{align}
Now, observing that from\,\eqref{metric1}
\begin{equation}
	(g^{(a)}(x))^{\frac38}=a^3\left(\widetilde g(x)\right)^{\frac38}\,,
\end{equation}
it is immediate to see that\,\eqref{effac1*} can be written as
\begin{align}\label{effac2*}
	e^{-\delta S_{\rm grav}}=\mathcal{A}\int\prod_{x}\left[\dd{\,\widehat{\overline\psi}(x)}\dd{\,\widehat{\psi}(x)}\right] e^{-\int\dd[4]x\,\sqrt{\,\widetilde g\,\,}\left[\frac12\left(\widehat{\overline\psi}(\widetilde{\slashed{\nabla}}\widehat\psi\,)-\overline{(\widetilde{\slashed{\nabla}}\widehat\psi\,)}\,\widehat\psi\right)+\,a\,m_\psi\,\widehat{\overline{\psi}}\widehat\psi\right]}\,,
\end{align}
where $\widetilde{\slashed{\nabla}}$ is the dimensionless Dirac operator $\widetilde{\slashed{\nabla}}\equiv a\slashed{\nabla}$, \,$\widehat\psi$ the dimensionless field defined as $\widehat\psi\equiv a^{\frac{3}{2}}\psi$, and $\mathcal A$ an $a$-independent factor. 

Expanding \,$\widehat\psi(x)$\, in terms of the eigenfunctions \cite{Camporesi:1995fb} of the operator $\widetilde{\slashed{\nabla}}$, we get ($\mathcal{C}$ is an $a$-independent term)
\begin{equation}\label{scalarres*}
	S_{\rm grav}^{\rm eff}\equiv S^{(a)}_{\rm grav}+\delta S_{\rm grav}=\frac{\pi\Lambda_{\rm cc}}{3G}a^4-\frac{2\pi}{G}a^2-\Tr\log\left(\widetilde{\slashed{\nabla}}+a\,m_\psi\right)+\mathcal C\,.
\end{equation}
Since
\begin{equation}
	\Tr\log\left(\widetilde{\slashed{\nabla}}+a\,m_\psi\right)=\frac12\Tr\log\left(\widetilde{\slashed{\nabla}}^{\,2}+a^2 m_\psi^2\right)\,,
\end{equation}
introducing the dimensionless proper-time $\tau$ with numerical cut $N\gg1$ (see\,\eqref{propertime*1}),  for $S^{\rm eff}_{\rm grav}$ we get
\begin{equation}\label{propertimefermion1}
	S_{\rm grav}^{\rm eff}=\frac{\pi\Lambda_{\rm cc}}{3G}a^4-\frac{2\pi}{G}a^2+\frac12\int_{1/N^2}^{+\infty}\frac{\dd{\tau}}{\tau}K^{(1/2)}(\tau)+\mathcal C\,,
\end{equation}
where
\begin{equation}
	{\rm K}^{(1/2)}(\tau)= \sum_{n=1/2}^{+\infty}D^{(1/2)}_n\, e^{-\tau\left(\lambda^{(1/2)}_n+a^2m_\psi^2\right)}
	\label{propertimefermion2}
\end{equation}
is the kernel of the operator ${\slashed{\nabla}}^{\,2}+a^2 m_\psi^2$. The eigenvalues $\lambda^{(1/2)}_n$ and the corresponding degeneracies $D^{(1/2)}_n$ of ${\slashed{\nabla}}^{\,2}$ are 
\begin{equation}
	\lambda^{(1/2)}_n=\left(n+\frac32\right)^2\quad ;\quad D^{(1/2)}_n=\frac{(2n+1)(2n+3)(2n+5)}{12}\quad,\quad\text{with}\,\,\,\, n=\frac12,\frac32,\frac52,\dots
	\label{eigenvalues*}
\end{equation}
Inserting\,\eqref{propertimefermion2} in\,\eqref{propertimefermion1}, performing the integration over $\tau$, summing over $n$ with the help of the EML formula, and finally expanding for $N\gg1$, we get
\begin{align}
	S_{\rm grav}^{\rm eff}&=\frac{\pi}{3}\left(\frac{\Lambda_{\rm cc}}{G}+\frac{m_\psi^4}{4\pi}\log N^2\right)a^4-2\pi\left[\frac{1}{G}+ \frac{m_\psi^2}{12\pi}\left(N^2-\log N^2\right)\right]a^2\nonumber\\
	&+\frac{N^4}{12}-\frac{N^2}{6}+\frac{11}{360}\log N^2+\mathcal{C}+\text{finite}\,.
	\label{effactionfermion}
\end{align} 

The similarity of\,\eqref{effactionfermion} with\,\eqref{effaction2} and\,\eqref{oneloopresHK} is evident. Again, the numerical cut $N$ is related to the physical cutoff $\Lambda_{\rm cut}$ ($\sim\mpl$) through the relation\,\eqref{Lambdapt}, and as for the previous cases the vacuum energy $\rho_{\rm vac}=\frac{\Lambda_{\rm cc}}{8\pi G}$ receives only a logarithmically divergent correction,
\begin{equation}\label{QCvefermion}
	\delta\left(\frac{\Lambda_{\rm cc}}{8\pi G}\right)=\frac{m_\psi^4}{32\pi^2}\log \frac{3\Lambda_{\rm cut}^2}{\Lambda_{\rm cc}}\,.
\end{equation}

As in sections \ref{oneloop} and \ref{scalar}, we observe that had we proceeded with the (incorrect) identification of $\Lambda_{\rm cut}$ through the relation $\Lambda_{\rm cut}=N/a$, spurious quartically and quadratically divergent terms would have appeared in the correction to $\frac{\Lambda_{\rm cc}}{G}$ (see\,\eqref{incorrectident2},\,\eqref{QCve} and comments below these equations). 
Comparing\,\eqref{QCvefermion} (fermions) with\,\eqref{QCve} (bosons) we see that both contributions to the vacuum energy are proportional to the forth power of the respective masses, although with opposite signs. Naturally, the total correction depends on the particle content of the full theory. Moreover, as already stressed at the end of section \ref{scalar} for the scalar case, considering the physical cutoff of order the Planck scale, and the masses of order the Fermi scale, the absence of quartic and quadratic divergences in both\,\eqref{QCvefermion} and\,\eqref{QCve} alleviates the cosmological constant problem of about $70$ orders of magnitude. 

It is important to stress that the fact that the vacuum energy has only a logarithmic sensitivity to the UV physical cutoff (see\,\eqref{onelooprespt},\,\eqref{QCve} and\,\eqref{QCvefermion}) comes from a careful treatment of the path integral that defines the effective action (in a curved spacetime background), and no supersymmetric embedding of the theory (supergravity) is needed to get rid of power-like divergences. These results represent a first step towards a more complete analysis of the cosmological constant problem. 

\section{Summary and conclusions}
\label{conclusions}

Considering pure gravity within the (euclidean) Einstein-Hilbert truncation, we performed the calculation of
the one-loop effective action $\Gamma^{1l}_{\rm grav}$ paying particular attention to important aspects that in the past were either missed or mistreated. In previous literature, this calculation was usually done resorting to proper-time regularization within the heat kernel expansion and gave rise to quartically and quadratically UV-sensitive contributions to the vacuum energy $\rho_{\rm vac}$. As a consequence, if the physical UV cutoff $\Lambda_{\rm cut}$ is $\sim\mpl$, a discrepancy of more that 120 orders of magnitude between the theoretical prediction and the measured value of $\rho_{\rm vac}$ is found. In this respect, it is worth to stress that up to now two main approaches have been adopted to dispose of these quartically and quadratically divergent terms, one formal, the other physical. The formal one consists in performing the calculation resorting to regularization schemes, such as dimensional regularization, that cancel power-like divergences by construction. Obviously, the application of these methods cannot be regarded as a physical way to solve the original problem: they simply implement a technical cancellation. On the more physical side, where the presence of these divergences is acknowledged, their cancellation is realized resorting to a supersymmetric embedding of the theory (supergravity). 

Our calculation shows that, if the measure in the path integral that defines $\Gamma^{1l}_{\rm grav}$ is appropriately taken into account and the physical cutoff $\Lambda_{\rm cut}$ is correctly introduced, $\rho_{\rm vac}$ presents only a (mild) logarithmic sensitivity \,to\, $\Lambda_{\rm cut}$. In other words, when the measure and the UV cutoff are correctly treated, the quartically and quadratically UV-sensitive contributions to $\rho_{\rm vac}$ are {\it automatically} absent. Our results are neither formal, nor resort to any ad hoc physical cancellation mechanism. They show that in $\rho_{\rm vac}$ quartic and quadratic divergences simply do not appear.

To test whether the absence of these power-like divergences persists even when matter contributions to $\rho_{\rm vac}$ are taken into account, we also  considered the free theory of a scalar and a Dirac field on a spherical gravitational background. 
We found that, if again the measure and the physical cutoff $\Lambda_{\rm cut}$ are correctly treated, also in these cases $\rho_{\rm vac}$  receives only logarithmically UV-sensitive contributions.
Since these contributions are proportional to the fourth power of masses, we are still left with a (at least) $50$ orders of magnitude discrepancy between the measured vacuum energy and the theoretical calculation. We think that the approach and the ideas put forward in the present work should be of help in the search for the (still elusive) physical mechanism that eventually drives $\rho_{\rm vac}$ down to its measured value, and hope to come back to this challenging issue in future work. 

In any case, the main result of this work, namely the proof that (irrespective of the theory being supersymmetric or not) the one-loop contribution to the vacuum energy does not contain power-like divergences, provides a new insight on the long standing cosmological constant problem.  

\section*{Acknowledgments}
We are particularly grateful to G. Veneziano for several insightful comments and discussions. We also thank N. Mavromatos, R. Percacci, A. Pilaftsis and D. Zappalà for useful discussions.	
The work of CB has been supported by the European Union – Next Generation EU through the research grant number P2022Z4P4B “SOPHYA - Sustainable Optimised PHYsics Algorithms: fundamental physics to build an advanced society” under the program PRIN 2022 PNRR of the Italian Ministero dell’Università e Ricerca (MUR). The work of VB, FC and AP is carried out within the INFN project  QGSKY.

\section*{Appendix A}
\label{appendixA}

In this Appendix we derive Eq.\,\eqref{result} in the text, closely following the strategy put forward in \cite{TaylorVeneziano}. Let us indicate with $h_n^{\mu\nu (i)}$ (transverse-traceless), $\xi_n^{\mu (i)}$ (transverse) and $\phi_n^{(i)}$ the pure spin-$2$, spin-$1$ and spin-$0$ eigenfunctions of the dimensionless Laplace-Beltrami operator $-\widetilde\square^{(s)}$ normalized as ($i$ is the degeneracy index)
{\footnotesize\begin{equation}
		\int\dd[4]{x}\sqrt{\,\widetilde g\,\,}\,h_n^{\mu\nu (i)}(x)h^{m (j)}_{\mu\nu}(x)=\int\dd[4]{x}\sqrt{\,\widetilde g\,\,}\,\xi_n^{\mu (i)}(x)\xi_\mu^{m (j)}(x)=\int\dd[4]{x}\sqrt{\,\widetilde g\,\,}\,\,\phi_n^{(i)}(x)\,\phi_m^{(j)}(x)=\delta^{ij}\delta_{nm}\,,
\end{equation}}

\noindent
and the corresponding eigenvalues with $\lambda_n^{(2)}$, $\lambda_n^{(1)}$ and $\lambda_n^{(0)}$ respectively (see\,\eqref{eigenvalues} in the text). The modes {\small $\{h_n^{\mu\nu},\,v_n^{\mu\nu},\,w_n^{\mu\nu},\,z_n^{\mu\nu}\}$}, where (from now on the degeneracy indexes are omitted)
\begin{align}
		v_n^{\mu\nu}&=\left[\frac12\left(\lambda_n^{(1)}-3\right)\right]^{-\frac12}\nabla^{(\mu}\xi_n^{\nu)}\,,\quad n=2,\dots\,,\nonumber\\
		w_n^{\mu\nu}&=\left[\lambda^{(0)}_n\left(\frac34\lambda^{(0)}_n-3\right)\right]^{-\frac12}\left(\nabla^{\mu}\nabla^\nu-\frac14 \widetilde g^{\,\mu\nu}\Box\right)\phi_n\,,\quad n=2,\dots\,,\nonumber\\
		z_n^{\mu\nu}&=\frac12\widetilde g^{\,\mu\nu}\phi_n\,,\quad n=0,1,2,\dots\,,
\end{align}

\noindent
form the orthonormal basis for symmetric tensors. Defining now the longitudinal vector modes
\begin{equation}
		l_n^{\mu}=\left(\lambda_n^{(0)}\right)^{-\frac12}\,\nabla^{\mu}\phi_n\,,\quad n=1,2,\dots\,,
\end{equation}
we observe that, together with the transverse modes $\xi^{\mu}_n$, they form the orthonormal basis for vectors.

Expanding the dimensionless graviton field $\widehat h^{\,\mu\nu}$ (see Eq.\,\eqref{hath} in the text) as
\begin{align}\label{expansionh}
		\widehat h^{\,\mu\nu}&=\sum_{n=2}^{\infty}a_n h^{\mu\nu}_n+\sum_{n=2}^{\infty}b_n v^{\mu\nu}_n+\sum_{n=2}^{\infty}c_n w^{\mu\nu}_n+\sum_{n=0}^{\infty}e_n z^{\mu\nu}_n\,,
\end{align}
the ghost field $\widehat v^{\,\mu}$ (Eq.\,\eqref{hat v} in the text) as
\begin{equation}\label{expansionv}
		\widehat v^{\,\mu}=\sum_{n=1}^{\infty}g_n\,\xi_n^\mu+\sum_{n=1}^{\infty}f_n\,l_n^\mu\,,
\end{equation}
and finally inserting\,\eqref{expansionh} and\,\eqref{expansionv} in\,\eqref{S2+Sgf} and\,\eqref{ghostaction2} respectively, we get
\begin{align}
		2\left(S_2+S_{\text{gf}}\right)&=\sum_{n=2}^{\infty}a_n^2\left[\lambda_n^{(2)}-2a^2\Lambda_{\rm cc}+8\right]+\sum_{n=2}^{\infty}b_n^2\left[\xi^{-1}\left(\lambda_n^{(1)}-3\right)-2a^2\Lambda_{\rm cc}+6\right]\nonumber\\
		&+\sum_{n=2}^{\infty}c_n^2\left[\xi^{-1}\left(\frac34\lambda_n^{(0)}-6\right)-\frac{\lambda_n^{(0)}}{2}-2a^2\Lambda_{\rm cc}+6\right]\nonumber\\
		&+\sum_{n=0}^{\infty}e_n^2\left[\frac{-3+\xi^{-1}}{2}\lambda_n^{(0)}+2a^2\Lambda_{\rm cc}\right]+\sum_{n=2}^{\infty}2e_n c_n(\xi^{-1}-1)\left[\lambda_n^{(0)}\left(\frac34 \lambda_n^{(0)}-3\right)\right]^{\frac{1}{2}}
\end{align}
and
\begin{align}
		 S_{\text{ghost}}&=\sum_{n=1}^{\infty}g_n^{*}\,g_n\left(\lambda_n^{(1)}-3\right)+\sum_{n=1}^{\infty}f_n^{*}\,f_n\left(\lambda_n^{(0)}-6\right)\,.\qquad\qquad
\end{align}
The functional measure\,\eqref{meas2} in the text can then be written as \cite{TaylorVeneziano}
\begin{equation}\label{measdecomp}
		\big[\mathcal{D}u(h)\mathcal{D}v_\rho^{*}\,\mathcal{D} v_\sigma\big]\sim\prod_{n=2}^{\infty}{\rm d}{a_n}\prod_{n=2}^{\infty}{\rm d}{b_n}\prod_{n=2}^{\infty}{\rm d}{c_n}\prod_{n=0}^{\infty}{\rm d}{e_n}\prod_{n=2}^{\infty}{\rm d}{g^*_n}\prod_{n=2}^{\infty}{\rm d}{g_n}\prod_{n=1}^{\infty}{\rm d}{f^*_n}\prod_{n=1}^{\infty}{\rm d}{f_n}\,.
\end{equation}

\noindent
Clearly there is no gaussian integration along the flat directions $g_1^*$ and $g_1$ (zero modes) of $S_{\rm ghost}$, and the related ghost fields are proportional to the Killing vectors $\xi^{\mu}_1$ (see \cite{TaylorVeneziano} for details). Finally, introducing\,\eqref{measdecomp} in\,\eqref{effacgrav1} and performing the gaussian integrations, Eq.\,\eqref{result} in the text is obtained.
Note that the term $\frac12\log\left(2a^2\Lambda_{\rm cc}\right)$ in\,\eqref{result} comes from the integration over $e_0$.

\section*{Appendix B}
\label{appendixB}

In this Appendix we derive Eq.\,\eqref{oneloopres} in the text. To this end, we report here Eq.\,\eqref{calculation1} that was obtained considering the direct product of eigenvalues
\begin{align}
	\delta S^{1l}_{\rm grav}=\frac12\sum_{n=2}^{N-2} &\Bigl[D_n^{(2)}\log\left(\lambda_n^{(2)}-2a^2\Lambda_{\rm cc}+8\right)+D_n^{(0)}\log\left(\lambda_n^{(0)}-2a^2\Lambda_{\rm cc}\right)\nonumber\\
	&-D_n^{(1)}\log\left(\lambda_n^{(1)}-3\right)-D_n^{(0)}\log\left(\lambda_n^{(0)}-6\right)\Bigr]+\frac12\log(2a^2\Lambda_{\rm cc})+\mathcal{B}\,.
	\label{calculation1*}
\end{align}
Using the identity
\begin{equation}\label{identity}
	\log\left(x/y\right)=-\int_{0}^{+\infty}\dd{u}\left[\left(x+u\right)^{-1}-\left(y+u\right)^{-1}\right]\,,
\end{equation}
Eq.\,\eqref{calculation1*} can be written as
\begin{align}
	\delta S^{1l}_{\rm grav}=-\frac12\int_{0}^{+\infty}\dd{u} \sum_{n=2}^{N-2} &\Bigl[\frac{D_n^{(2)}}{u+\lambda_n^{(2)}-2a^2\Lambda_{\rm cc}+8}+\frac{D_n^{(0)}}{u+\lambda_n^{(0)}-2a^2\Lambda_{\rm cc}}-\frac{D_n^{(1)}}{u+\lambda_n^{(1)}-3}-\frac{D_n^{(0)}}{u+\lambda_n^{(0)}-6}\nonumber\\
	&-\frac{D_n^{(2)}-D_n^{(1)}}{u+1}\Bigr]+\frac12\log(2a^2\Lambda_{\rm cc})+\mathcal{B}\,.
	\label{calculation1**}
\end{align}
Performing now the sum over $n$, the integration over $u$, and finally expanding the result for $N\gg1$, we obtain Eq.\,\eqref{oneloopres} in the text for $\delta S_{\rm grav}^{1l}$, with $\mathcal{F}(a^2\Lambda_{\rm cc})$ given by {\begin{align}
		&\mathcal{F}(a^2\Lambda_{\rm cc})=9\,a^2\Lambda_{\rm cc}-\frac{ a^2\Lambda_{\rm cc}  \sqrt{8 a^2\Lambda_{\rm cc}+9}}{6}\,\log\Gamma \left(\frac{7-\sqrt{8 a^2\Lambda_{\rm cc}+9}}{2}\right) -5\,a^2\Lambda_{\rm cc}\,\psi ^{(-2)}\left(\frac{7+\sqrt{8 a^2 \Lambda_{\rm cc} -15}}{2} \right) \nonumber\\
		&-5\,a^2\Lambda_{\rm cc}\,\psi ^{(-2)}\left(\frac{7-\sqrt{8 a^2 \Lambda_{\rm cc} -15}}{2}\right) -a^2\Lambda_{\rm cc} \,\psi ^{(-2)}\left(\frac{7+\sqrt{8 a^2\Lambda_{\rm cc}+9}}{2} \right)\nonumber\\
		&-a^2\Lambda_{\rm cc}\,\psi ^{(-2)}\left(\frac{7-\sqrt{8 a^2\Lambda_{\rm cc}+9}}{2}\right)+\frac{a^2\Lambda_{\rm cc}  \,\sqrt{8 a^2\Lambda_{\rm cc}+9}}{6} \,\log\Gamma\left(\frac{7+\sqrt{8 a^2\Lambda_{\rm cc}+9}}{2} \right)-5 \log (120) \nonumber
		\end{align}
		\begin{align}
		&+\frac{49 \log (A)}{3}-2 \sqrt{\frac{11}{3}} \log\Gamma\left(\frac{7+\sqrt{33}}{2} \right)-\frac{5\left(a^2 \Lambda_{\rm cc} -5\right) \sqrt{8 a^2 \Lambda_{\rm cc} -15}}{6}\,\log\Gamma\left(\frac{7-\sqrt{8 a^2 \Lambda_{\rm cc} -15}}{2}\right)\nonumber\\
		&-\frac{\sqrt{8 a^2\Lambda_{\rm cc}+9}}{6}\,\log\Gamma\left(\frac{7-\sqrt{8 a^2\Lambda_{\rm cc}+9}}{2}\right)+3 \left(\psi ^{(-4)}(1)+\psi ^{(-4)}(6)\right)+\psi ^{(-4)}\left(\frac{7-\sqrt{33}}{2}\right)\nonumber\\
		&+\psi ^{(-4)}\left(\frac{7+\sqrt{33}}{2}\right)-5 \left(\psi ^{(-4)}\left(\frac{7+\sqrt{8 a^2 \Lambda_{\rm cc} -15}}{2} \right)+\psi ^{(-4)}\left(\frac{7-\sqrt{8 a^2 \Lambda_{\rm cc} -15}}{2}\right)\right)\nonumber\\
		&-\psi ^{(-4)}\left(\frac{7+\sqrt{8 a^2\Lambda_{\rm cc}+9}}{2}\right)-\psi ^{(-4)}\left(\frac{7-\sqrt{8 a^2\Lambda_{\rm cc}+9}}{2}\right)+\frac{15}{2}\left( \psi ^{(-3)}(1)-\psi ^{(-3)}(6)\right)\nonumber\\
		&-\frac{\sqrt{33}}{2} \,\psi ^{(-3)}\left(\frac{7+\sqrt{33}}{2}\right)-\frac{5\sqrt{8 a^2 \Lambda_{\rm cc} -15}}{2}\, \psi ^{(-3)}\left(\frac{7-\sqrt{8 a^2 \Lambda_{\rm cc} -15}}{2}\right)\nonumber\\
        &-\frac{\sqrt{8 a^2\Lambda_{\rm cc}+9}}{2} \,\psi ^{(-3)}\left(\frac{7-\sqrt{8 a^2\Lambda_{\rm cc}+9}}{2}\right)+\frac{33}{4}\left(\psi ^{(-2)}(1)+\psi ^{(-2)}(6)\right)\nonumber\\
		&+\frac{49}{12}\left(\psi ^{(-2)}\left(\frac{7-\sqrt{33}}{2}\right)+\psi ^{(-2)}\left(\frac{7+\sqrt{33}}{2}\right)\right)\nonumber\\
		&+\frac{175}{12}\left(\psi ^{(-2)}\left(\frac{7+\sqrt{8 a^2 \Lambda_{\rm cc} -15}}{2}\right)+\psi ^{(-2)}\left(\frac{7-\sqrt{8 a^2 \Lambda_{\rm cc} -15}}{2}\right)\right)\nonumber\\
		&-\frac{13}{12}\left(\psi ^{(-2)}\left(\frac{7+\sqrt{8 a^2\Lambda_{\rm cc}+9}}{2}\right)+\psi ^{(-2)}\left(\frac{7-\sqrt{8 a^2\Lambda_{\rm cc}+9}}{2}\right)\right)+\frac{\sqrt{33}}{2}\,\psi ^{(-3)}\left(\frac{7-\sqrt{33}}{2}\right)\nonumber\\
		&+2\,\sqrt{\frac{11}{3}}\,\log\Gamma\left(\frac{7-\sqrt{33}}{2}\right)+\frac{5\left(a^2 \Lambda_{\rm cc} -5\right)\sqrt{8 a^2 \Lambda_{\rm cc} -15}}{6}\,\log\Gamma \left(\frac{7+\sqrt{8 a^2 \Lambda_{\rm cc} -15}}{2}\right)\nonumber\\
		&+\frac{5\sqrt{8 a^2 \Lambda_{\rm cc} -15}}{2}\,\psi ^{(-3)}\left(\frac{7+\sqrt{8 a^2 \Lambda_{\rm cc} -15}}{2}\right)+\frac{\sqrt{8 a^2\Lambda_{\rm cc}+9}}{6}\, \log\Gamma\left(\frac{7+\sqrt{8 a^2\Lambda_{\rm cc}+9}}{2}\right) \nonumber\\
		&+\frac{\sqrt{8 a^2\Lambda_{\rm cc}+9}}{2}\,\psi ^{(-3)}\left(\frac{7+\sqrt{8 a^2\Lambda_{\rm cc}+9}}{2}\right)+\frac{7 \zeta (3)}{4 \pi ^2}-\frac{2\,\zeta '(-3)}{3}-\frac{20801}{1080}\,,
	\end{align}}

\noindent
where $A$ is the Glaisher's constant ($A\simeq1.282427$), $\zeta(z)$ is the Riemann zeta function ($\zeta (3)\simeq1.20206$ and $\zeta '(-3)\simeq0.00538$), and $\psi ^{(-n)}(z)$ (with $n$ positive integer) are the polygamma functions of negative order defined as\,\cite{Adamchik}
\begin{equation}
	\psi ^{(-n)}\left(z\right)=\frac{1}{(n-2)!}\int_{0}^z\dd{t}(z-t)^{n-2} \log\,\Gamma\left(t\right)\qquad\qquad \text{for}\,\,\Re(z)>0\,.
\end{equation}


\begin{thebibliography}{100}
	
	\vskip 10pt
	\bibitem{perl} S.Perlmutter {\it et al.}, Nature {\bf 391}, 51 (1998);
	S. Perlmutter {\it et al.}, APJ {\bf 517}, 565 (1999); 
	P.M. Garnavich {\it et al.}, APJ Lett.{\bf 493}, L53 (1998);
	A.G. Riess {\it et al.}, Astron. J.{\bf 116}, 1009 (1998). 
	
	\bibitem{ParticleDataGroup:2024cfk}
	S.~Navas \textit{et al.} [Particle Data Group],
	\emph{Review of particle physics},
	Phys. Rev. D \textbf{110} (2024) no.3, 030001.

	\bibitem{Schwinger:1951nm}
	J.~S.~Schwinger,
	\emph{On gauge invariance and vacuum polarization},
	Phys. Rev. \textbf{82} (1951), 664.
	
	\bibitem{DeWitt:1964mxt}
	B.~S.~DeWitt,
	\emph{Dynamical theory of groups and fields}. In: Relativity, Groups and Topology, ed. B.S. DeWitt and C. DeWitt,
	Conf. Proc. C \textbf{630701} (1964), 585.
	
	\bibitem{Seeley1}
	R.~T.~Seeley,
	\emph{Singular integrals and boundary value problems}, Amer. J. Math. 88 (1966) 781.

	\bibitem{Seeley2}
	R.~T.~Seeley,
	\emph{The resolvent of an elliptic boundary value problem}, Amer. J. Math. 91 (1969) 889.
	
	\bibitem{DeWitt:1975ys}
	B.~S.~DeWitt,
	\emph{Quantum Field Theory in Curved Space-Time},
	Phys. Rept. \textbf{19} (1975), 295.
	
	\bibitem{FradkinTseytlin}
	E.~S.~Fradkin and A.~A.~Tseytlin,
	\emph{On the New Definition of Off-shell Effective Action},
	Nucl. Phys. B \textbf{234} (1984), 509.
	
	\bibitem{TaylorVeneziano}
	T.~R.~Taylor and G.~Veneziano,
	\emph{Quantum Gravity at Large Distances and the Cosmological Constant},
	Nucl. Phys. B \textbf{345} (1990), 210.
	
	\bibitem{Branchina:2022jqc}
	C.~Branchina, V.~Branchina, F.~Contino and N.~Darvishi,
	\emph{Dimensional regularization, Wilsonian RG, and the naturalness and hierarchy problem},
	Phys. Rev. D \textbf{106} (2022) no.6, 065007.
	
	\bibitem{Branchina:2023ogv}
	C.~Branchina, V.~Branchina, F.~Contino and A.~Pernace,
	\emph{Does the Cosmological Constant really indicate the existence of a Dark Dimension?}, Int. Jour. of Geom. Meth. in Mod. Phys. (doi:10.1142/S0219887824503055).
	
	\bibitem{Branchina:2024ljd}
	C.~Branchina, V.~Branchina, F.~Contino and A.~Pernace,
	\emph{Dark Dimension and the Effective Field Theory limit}, Int. Jour. of Geom. Meth. in Mod. Phys. (doi:10.1142/S0219887824503031).
	
	\bibitem{Branchina:2023rgi}
	C.~Branchina, V.~Branchina and F.~Contino,
	\emph{Naturalness and UV sensitivity in Kaluza-Klein theories},
	Phys. Rev. D \textbf{108} (2023) no.4, 045007.
	
	\bibitem{Fradkin:1976xa}
	E.~S.~Fradkin and G.~A.~Vilkovisky,
	\emph{On Renormalization of Quantum Field Theory in Curved Space-Time},
	Lett. Nuovo Cim. \textbf{19} (1977), 47.
	
	\bibitem{Donoghue:2020hoh}
	J.~F.~Donoghue, \emph{Cosmological constant and the use of cutoffs},
	Phys. Rev. D \textbf{104} (2021) no.4, 045005.
	
	\bibitem{Vilkovisky:1984st}
	G.~A.~Vilkovisky, \emph{The Unique Effective Action in Quantum Field Theory},
	Nucl. Phys. B \textbf{234} (1984), 125.
	
	\bibitem{DeWitt:1987te}
	B.S.~DeWitt, \emph{The effective action}, in:\,Architecture of Fundamental Interactions at Short Distances:\,\,Proceedings, Les Houches 44th Summer School of Theoretical Physics, France, July 1-August 8, 1985, pt.2, Eds. P.~Ramond and R.~Stora, p. 1023.
	
	\bibitem{Abbott:1980hw}
	L.~F.~Abbott,
	\emph{The Background Field Method Beyond One Loop},
	Nucl. Phys. B \textbf{185} (1981), 189.
	
	\bibitem{Abbott:1981ke}
	L.~F.~Abbott,
	\emph{Introduction to the Background Field Method},
	Acta Phys. Polon. B \textbf{13} (1982), 33
	CERN-TH-3113.
	
	\bibitem{Fradkin:1973wke}
	E.~S.~Fradkin and G.~A.~Vilkovisky,
	\emph{S matrix for gravitational field. ii. local measure, general relations, elements of renormalization theory},
	Phys. Rev. D \textbf{8} (1973), 4241.
	
	\bibitem{Unz:1985wq}
	R.~K.~Unz,
    \emph{Path Integration and the Functional Measure},
	Nuovo Cim. A \textbf{92} (1986), 397.
	
	
	\bibitem{Honerkamp:1971xtx}
	J.~Honerkamp and K.~Meetz,
	\emph{Chiral-invariant perturbation theory},
	Phys. Rev. D \textbf{3} (1971), 1996.
	
	\bibitem{Becker:2020mjl}
	M.~Becker and M.~Reuter,
	\emph{Background Independent Field Quantization with Sequences of Gravity-Coupled Approximants},
	Phys. Rev. D \textbf{102} (2020) no.12, 125001.

	\bibitem{Becker:2021pwo}
	M.~Becker and M.~Reuter,
	\emph{Background independent field quantization with sequences of gravity-coupled approximants. II. Metric fluctuations},
	Phys. Rev. D \textbf{104} (2021) no.12, 125008.
	
	\bibitem{Ferrero:2024yvw}
	R.~Ferrero and R.~Percacci,
	\emph{The cosmological constant problem and the effective potential of a gravity-coupled scalar},
	JHEP \textbf{09} (2024), 074.
		
	\bibitem{noi}
	C.~Branchina, V.~Branchina, F.~Contino, R.~Gandolfo and A.~Pernace,
	\emph{work in progress}.
	
	\bibitem{interacting}
	C.~Branchina, V.~Branchina, F.~Contino, R.~Gandolfo and A.~Pernace,
	\emph{work in progress}.
	
	\bibitem{Camporesi:1995fb}
	R.~Camporesi and A.~Higuchi,
	\emph{On the Eigenfunctions of the Dirac operator on spheres and real hyperbolic spaces},
	J. Geom. Phys. \textbf{20} (1996), 1.
	
	\bibitem{Adamchik}
	V.~S.~Adamchik,
	\emph{Polygamma functions of negative order},
	Journal of Computational and Applied Mathematics, \textbf{100} (1998) no.2, 191.
	
\end{thebibliography}
\end{document}